\documentclass{article}
\usepackage{graphicx}  % standard LaTeX graphics tool;
\usepackage{amsmath}   % For getting proper math eqns
\usepackage{cite}
\usepackage{amssymb}   % Used for getting various symbols eg. gtrsim, lesssim etc.
\usepackage{bm} % For bold math (esp of lower greek letters)
\usepackage{dcolumn}% Align table columns on decimal point
\usepackage{color}
\usepackage{mathrsfs}
\usepackage{amsfonts}
\usepackage{varioref}
\usepackage{braket}
\usepackage{verbatim}
\usepackage{subcaption}

\RequirePackage[colorlinks,citecolor=blue,urlcolor=magenta,linkcolor=blue]{hyperref}
\labelformat{section}{Section #1} 
\labelformat{subsection}{Section #1} 
\labelformat{subsubsection}{Section #1}
\labelformat{subsubsubsection}{Section #1}
\labelformat{equation}{Eq.~(#1)} 
\labelformat{figure}{Fig.~#1} 
\labelformat{subfigure}{Fig.~\thefigure#1} 
\labelformat{table}{Tab.~#1} 
\labelformat{appendix}{Appendix #1}
\title{Complex time route to quantum backreaction}

\author{Karthik Rajeev\footnote{karthik@iucaa.in},~ 
	\\
	{\small{IUCAA, Post Bag 4, Ganeshkhind, Pune University Campus, Pune 411007, India}}}
\date{}    

\begin{document}
	\maketitle
	\begin{abstract}
		We consider the backreaction of a quantum system $q$ on an effectively classical degree of freedom $C$ that is interacting with it. The backreaction equation based on the standard path integral formalism gives the so-called `in-out' backreaction equation, which has several serious pathologies. One can use a different backreaction prescription, referred to as the `in-in' approach, which resolves all the issues of `in-out' backreaction equation. However, this procedure is usually invoked in a rather ad hoc manner. Here we provide a rigorous path integral derivation of the `in-in' approach by exploiting the concept of quantum evolution along complex time contours. It turns out that, this approach can also be used to study both the `in-in' and `in-out' backreaction equations in a unified manner. 
	\end{abstract}    
	
	\section{Introduction}    
	
	The probability amplitude $A(Q_f,t;Q_i,t_i)$ that a system, which was initially, say at $t_i$, in the configuration $Q_i$ may be found in the configuration $Q_{f}$ at a later time $t$ is given by \cite{feynman2010quantum}
	\begin{align}
	A(Q_f,t;Q_i,t_i)=\int_{Q(t_i)=Q_i}^{Q(t_f)=Q_f} \mathcal{D}[Q]\,\,e^{i\frac{S[Q]}{\hbar}}
	\end{align}
	where $\mathcal{D}[Q]$ is an appropriate functional measure. In the classical limit, defined by $\hbar\rightarrow 0$, the stationary phase approximation can be invoked to show that the dominant contribution to this integral comes from the configurations that satisfy $\delta S/\delta Q=0$. 
	
	When the degrees of freedom of a system can be naturally divided into two subsystems, say $C$ and $q$, apart from the classical limit (viz. the $\hbar\rightarrow 0$ limit), one can also study another useful limit. This corresponds to the limit in which one subsystem, say $C$, is effectively classical while the other is quantum mechanical. There are numerous physical systems in which such a limit arises in a natural manner, like for e.g., in the study of quantum field theory (QFT) in curved spacetime. In the study of such systems, quantum backreaction refers to the correction to classical dynamics of the subsystem $C$ due to the feedback from the quantum excitations of $q$. 
	
	To explore this in  some more detail, let us consider a $C-q$ system described by the following action:
	\begin{align}
	S[q,C]=S_1[q]+S_2[C]+S_{12}[q,C]
	\end{align}
	The first two terms, namely, $S_1[q]$ and $S_2[C]$, represent the free evolution of the subsystems $q$ and $C$, respectively. The interaction between the subsystems is described by $S_{12}[q,C]$. We shall now assume that there exists a limit in which the subsystem $C$ is effectively classical, while $q$ is quantum mechanical. One can then study this limit of the $C-q$ system at two `levels'. At level-I, we ignore the backreaction of $q$ on $C$. We then deal with the quantum dynamics of $q$ while assuming that the classical subsystem is described by a given configuration $C(t)$. The kernel $A(q_f,t;q_i,t)$ of the subsystem $q$, at this level, is then given by:
	\begin{align}
	A(q_f,t;q_i,t_i)=\int_{q(t_i)=q_i}^{q(t_f)=q_f} \mathcal{D}[q]\,\,e^{i\frac{S[q,C(t)]}{\hbar}}
	\end{align}    
	Thus, the level-I describes \textit{quantum theory in a classical background}. At the next level, namely level-II, we want to take into account the effects of quantum fluctuations of the subsystem $q$ on $C$ by an effective classical description. The corresponding equation of motion for $C$, including the backreaction, is then expected to take the following general form:
	\begin{align}\label{level2}
	\frac{\delta S_2[C]}{\delta C}+\left\langle\frac{\delta S_{12}[q,C]}{\delta C}\right\rangle=0
	\end{align} 
	where, $\left\langle\,\,\,\right\rangle$ denotes a suitable operation to construct a c-number from the quantum theory of $q$. While level-I is relatively well understood, there are fundamental issues at the Level-II. One of the major issues stems from the fact that there is no general procedure to \textit{derive} the second term of \ref{level2} in a systematic manner. We will now elaborate on these issues.
	
	One approach towards the backreaction equation, that is often discussed in the literature, uses an effective action $S_{eff}[C]$ for the system $C$. It seems natural to define this effective action by `integrating out' the quantum degree of freedom $q$ in the following manner (see for e.g., \cite{dewitt2003global,padmanabhan2016quantum}):
	\begin{align}\label{seffdef}
	\exp\left(\frac{i}{\hbar}S_{eff}[C]\right)\equiv\int \mathcal{D}[q]\exp\left(\frac{i}{\hbar}S[q,C]\right)
	\end{align}    
	To obtain the explicit dynamical equation that describes the backreaction on the system $C$, we may demand that $\delta\textrm{Re}[S_{eff}]/\delta C=0$ for the effective classical `trajectory' $C(t)$. The justification for this demand is that the contribution to path integral of $\exp{iS_{eff}[C]/\hbar}$ over all configuration of $C$ is dominated by configurations in the neighbourhood of those `trajectories' that satisfy $\delta\textrm{Re}[S_{eff}]/\delta C=0$. The backreaction equation for $C$ that follows from this prescription can be shown to be given by:
	\begin{align}\label{breqn1}
	\frac{\delta S_2[C]}{\delta C}+\textrm{Re}\left[\frac{1}{\braket{\textrm{out}|\textrm{in}}}\braket{\textrm{out}|\left(\frac{\delta S_{12}[q,C]}{\delta C}\right)|\textrm{in}}\right]=0
	\end{align}
	where, $\ket{\textrm{out}}$ and $\ket{\textrm{in}}$ are the appropriate vacuum states at, respectively, the asymptotic future and past of the $q-$subsystem in the background of $C(t)$. The backreaction equation obtained from varying the effective action is therefore equivalent to choosing the operation $\braket{\,\,\,}$ in \ref{level2} to be $\textrm{Re}[\braket{\textrm{out}|(\,\,\,)|\textrm{in}}/\braket{\textrm{out}|\textrm{in}}]$. Hence, this prescription to backreaction is referred to as the `in-out' approach.  
	
	Unfortunately, there are some severe issues in this approach. First, the presence of $\ket{\textrm{out}}$ in the definition of `in-out' approach implies that the corresponding backreaction equation is non-causal. Second, the dynamics of $C$ obtained by this approach does not seem to completely incorporate the effects of particle production (see \ref{inout} for details). More specifically, the energy conservation equation that follows from \ref{breqn1} does not have the correct contribution expected from the pair creation process. These undesirable features cannot be completely resolved within the `in-out' formalism.     
	
	This motivates the natural question: How can one create a better prescription that will remedy these issues? We could make a reasonable conjecture that the correct backreaction equation corresponds to the one in which the operation $\braket{\,\,\,}$ in \ref{level2} is given by $\braket{\textrm{in}|(\,\,\,)|\textrm{in}}$, i.e., just the expectation value evaluated with respect to the `in-vacuum' state. The explicit form of the backreaction equation is then given by:
	\begin{align}\label{breqn2}
	\frac{\delta S_2[C]}{\delta C}+ \left\langle\textrm{in}\left|\frac{\delta S_{12}[q,C]}{\delta C}\right|\textrm{in}\right\rangle=0
	\end{align}
	This prescription, which we shall refer to as the `in-in' approach, is supported by the fact that the energy conservation equation that follows from \ref{breqn2} has the correct form, as for example discussed in \cite{Rajeev:2017uwk}. Moreover, causality is also retained in this approach. The main drawback concerning the `in-in' prescription is that the manner in which we have postulated -- rather than derived -- \ref{breqn2}. An attempt to derive the backreaction equation from the standard path integral approach seems to only give us \ref{breqn1}, i.e., the `in-out' backreaction equation. 
	
	There is, though, a different path integral approach that is expected to give the `in-in' backreaction equation in the appropriate limit. This corresponds to the Schwinger-Keldysh formalism\cite{Schwinger:1960qe,Keldysh:1964ud,Feynman:1963fq}, a path integral based approach adapted to address non-equilibrium quantum systems, which naturally contains a prescription to generate `in-in' expectation values of operators. To implement this method, however, one has to first formulate path integral over a configuration space of the variables $\bar{q}$ and $\bar{C}$ obtained by doubling the degrees of freedom of $q$ and $C$, respectively, i.e., $\bar{q}\equiv\{q_+,q_-\}$ and $\bar{C}\equiv\{C_{+},C_{-}\}$. This `doubling' is again rather ad hoc and hence not quite satisfactory. 
	
	Can we provide a more natural derivation of the `in-in' backreaction directly from path integral formalism? In fact, we can, and the main motivation of this paper is to provide such a derivation for a specific class of model $C-q$ systems that has broad applications in physics. 
	
	In order to do this, we first describe an approach to studying the evolution of a quantum system along a complex time-contour. Then, for a specific $C-q$ system, we describe how one can arrive at the explicit form of the effective action $S^{\mathcal{T}}_{eff}[C]$ for time evolution along an arbitrary time contour $\mathcal{T}$. Next, we introduce two specific contours $\mathcal{T}_1$ and $\mathcal{T}_2$, shown in \ref{fig:1} and \ref{fig:2}. We then show that when the contour is chosen to be $\mathcal{T}_1$, the effective equation of motion of $C$ that follows from $\delta S^{\mathcal{T}_1}_{eff}[C]/\delta C=0$ corresponds to that of the `in-out' approach. On the other hand, when the contour is chosen to be $\mathcal{T}_2$, the effective classical equation of motion that follows from $\delta S^{\mathcal{T}_2}_{eff}[C]/\delta C=0$ is precisely the `in-in' backreaction equation. Thus, the concept of time evolution along complex time-contours offers a unified approach to get both the `in-out' as well as the `in-in' backreaction equations. For reasons discussed earlier, $\mathcal{T}_2$ is the contour appropriate for the study of causal evolution of the effectively classical variable $C$, with all the effects of pair creation process also correctly taken into account. (Hereafter, we work in a system of units with $\hbar=1$.)

	\section{A useful model $C-q$ system}\label{setup}
	In this work, we will illustrate the ideas for a $C-q$ system described by the following Lagrangian. 
	\begin{align}\label{Cq_lag}
	\mathcal{L}=\frac{m(C)}{2} \left[\dot{q}^2-\omega^2(C)q^2\right]+M\left[\frac{\dot{C}^2}{2}-V(C)\right]
	\end{align}
	For a given background configuration of $C(t)$, the $q$ system is described by a time dependent harmonic oscillator(TDHO) of mass $m(C(t))$ and frequency $\omega(C(t))$. This feature of the $q$ system is shared by the Fourier modes of many quantum fields interacting with a classical background\cite{Parker:1968mv,Mahajan:2007qg}. To see this in a specific example, consider the action for the system consisting of the scalar field $\Phi$ and the scale factor $a$ of the Friedman universe with the metric $ds^2=-dt^2+a^2(t)|d\mathbf{x}|^2$ in the minisuperspace model \cite{Vilenkin:1994rn}. This is essentially given by the scalar field action plus the Einstein-Hilbert action, written as a functional of the scale factor. After some simplifications and introducing the variable $\xi=a^{3/2}$, the action takes the form:
	\begin{align}\label{superspaceaction}
	S[a,\left\{\Phi_{\mathbf{k}}\right\}]=V\int dt\left(-\frac{8}{3}\dot{\xi}^2\right)+\sum_{\mathbf{k}}\int dt\left[\frac{1}{2}|\dot{\Phi}_k|^2-\frac{1}{2}\left(\mu^2+\frac{k^2}{\xi^{4/3}}\right)|\Phi_{k}|^2\right]
	\end{align}   
	Comparing \ref{Cq_lag} and \ref{superspaceaction}, it is easy to make the following identification: $\xi=C$, $M=-8/3$, $V(C)= 0$, $m(C)=1$, $\omega^2(C)=(\mu^2+k^2/\xi^{4/3})$ and each Fourier mode, labelled by $\mathbf{k}$, can be identified with $q$. Another example, in which the study of our model $C-q$ system can shed some light, corresponds to a complex scalar field $\Psi$ interacting with a homogeneous electric field background in flat spacetime, say, along the $x-$axis. Such an electric field configuration can be described by the vector potential $A_{i}=(0,A(t),0,0)$. The corresponding action takes the following form:
	\begin{align}\label{superspaceactionA}
	S[A,\left\{\Psi_{\mathbf{k}}\right\}]=V\int dt \left(\frac{1}{2}\dot{A}^2\right) +\sum_{\mathbf{k}}\int dt \left[|\dot{\Psi}_{\mathbf{k}}|^2-\left(\mu^2+|\mathbf{k}_{\perp}|^2+\left\{k_x+qA(t)\right\}^2\right)|\Psi_{\mathbf{k}}|^2\right]
	\end{align}    
	where, $\mathbf{k}_{\perp}=(0,k_y,k_z)$. In this case, a comparison with \ref{Cq_lag} shows the following identification: $A=C$, $M=1$, $V(C)=0$,  $m(C)=2$, $\omega^2(C)=\left(\mu^2+|\mathbf{k}_{\perp}|^2+\left\{k_x+qA(t)\right\}^2\right)$ and each Fourier mode of $\Psi$, labelled by $\mathbf{k}$, can be identified with $q$.   
	
	Though there is an infinite number of oscillators in both \ref{superspaceaction} and \ref{superspaceactionA}, corresponding to, respectively, the Fourier modes the scalar fields $\Phi$ and $\Psi$, they are all mutually decoupled. Therefore, to understand the backreaction effects on, say $\xi$, we may start by considering the effects of only one oscillator and the results obtained in that case can easily be generalized to the case of a collection of mutually decoupled oscillators, each coupled to $\xi$. A similar argument also holds for the case of backreaction on the vector potential $A(t)$. This is the primary motivation for our choice of the Lagrangian in \ref{Cq_lag}. The $q$-independent part of $\mathcal{L}$, namely, the one describing the free evolution of $C$, has been chosen to be of a simple form for convenience and our analysis can be easily extended to any arbitrary form of this part.\footnote{The existence of an infinite number of degrees of freedom, of course, introduces several extra complications which are \textit{not} present in a \textit{finite} dimensional system. These issues manifest as the divergences in QFT. Such issues are usually resolved through careful regularization schemes and renormalization techniques. We will not discuss these issues here, since they are not directly relevant to our work.}
	
	It is clear that to study the semi-classical aspects of the system defined by \ref{Cq_lag} we need to understand the quantum dynamics of a TDHO. Since this is a fairly well-studied subject, we will only quote the results relevant for this work and delegate the details and derivations to the Appendix.   
	\subsection{Effective action from the standard path integral }\label{inout}
\begin{figure}[h!]
		\centering
		\begin{subfigure}[b]{0.5\linewidth}
			\includegraphics[scale=.28]{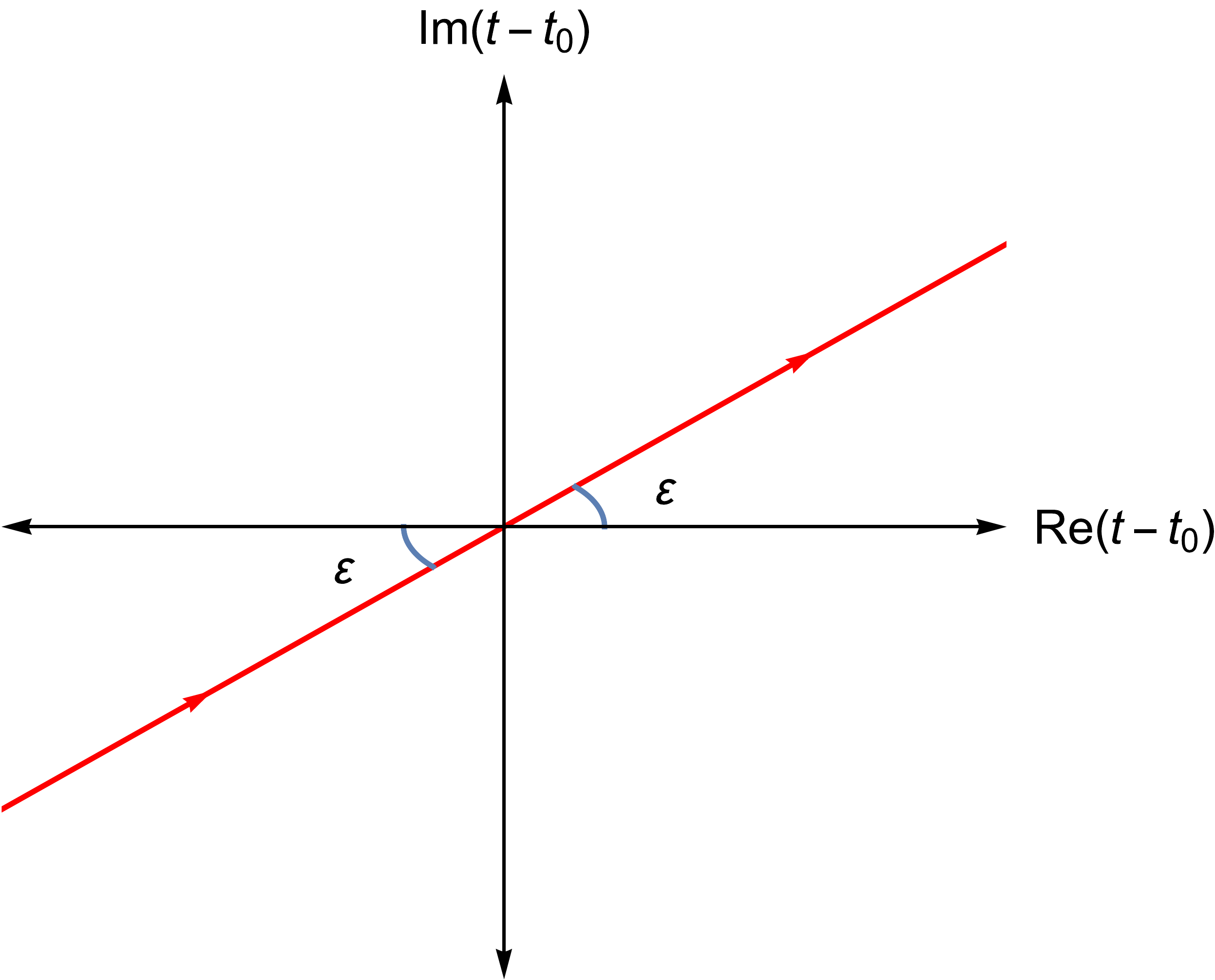}
			\caption{$\mathcal{T}_1$: the natural complex time contour that is relevant in the `in-out' formalism.}
			\label{fig:1}
		\end{subfigure}\vspace{5mm}
		\begin{subfigure}[b]{0.5\linewidth}
			\includegraphics[scale=.28]{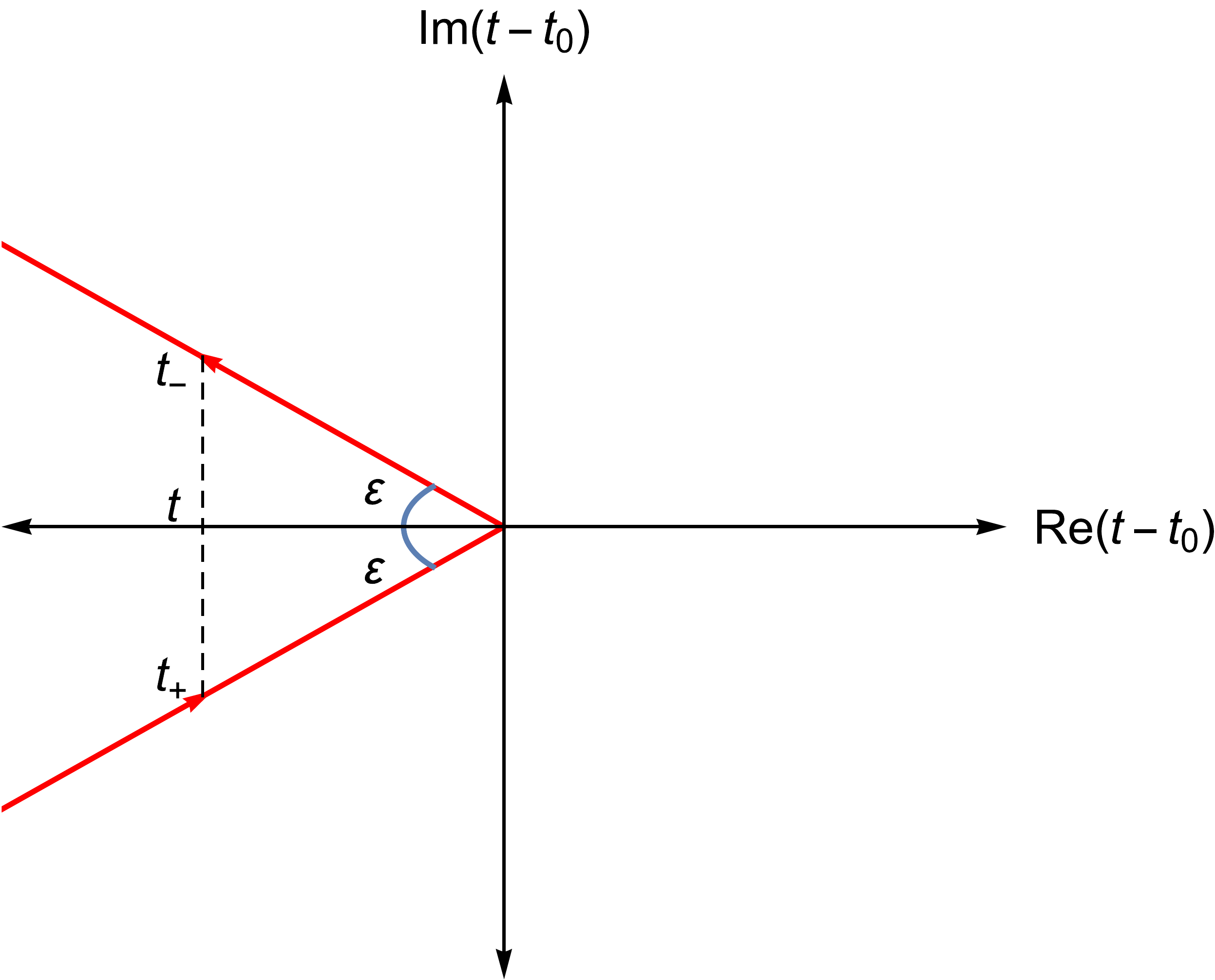}
			\caption{$\mathcal{T}_2$: the natural complex time contour that is relevant in the `in-in' formalism.}
			\label{fig:2}
		\end{subfigure}
		\caption{Different contours used for deriving the backreaction equations.}
		\label{figure}
	\end{figure}

	Before going into the derivation of `in-in' backreaction equation, we shall first briefly review the standard `in-out' approach. For this purpose, we start by evaluating the effective action $S_{eff}[C]$, obtained by `integrating out' the $q$ degree of freedom, as shown in  \ref{seffdef}. For our model $C-q$ system, the definition of $S_{eff}[C]$ takes the following form:
	\begin{align}\label{ioseff}
	\exp\left(i S_{eff}[C]\right)\equiv \exp\left(i\int_{-\infty}^{\infty} dt\, M\left[\frac{\dot{C}^2}{2}-V(C)\right]\right)\,\,\int \mathcal{D}[q]\exp\left(-\frac{i}{2}\int_{-\infty}^{\infty}dt\,\, q \hat{O}[C] q \right)
	\end{align} 
	where,
	\begin{align}
	\hat{O}[C]=\frac{d}{dt}\left(m(C)\frac{d}{dt}\right)+m(C)\omega^2(C)
	\end{align}
	There is, however a well-known issue here, namely that, the Gaussian path integral in \ref{ioseff}, strictly speaking, does not converge. One way of making sense of this path integral is to first deform the range of $t$ in the integral $\int_{-\infty}^{\infty}dt\,\, q \hat{O}[C] q$ from the real axis to the contour $\mathcal{T}_1$ shown in \ref{fig:1}. This corresponds to the $i\epsilon-$prescription in standard path integral approach to QFT. The path integral in \ref{ioseff} is replaced by the following factor:
	\begin{align}\label{ioseff2}
	I= \int\mathcal{D}[q]\exp\left(-\frac{i}{2}\int_{\mathcal{T}_1}dt\,\, q \hat{O}[C] q \right)
	\end{align}  
	This Gaussian path integral can be explicitly evaluated to get the following final form for $S_{eff}[C]$:
	
	\begin{align}\label{actionforCq}
	S_{eff}[C]=M\int dt \left[\frac{\dot{C}^2}{2}-V(C)\right]+\frac{i}{2}\log[\textrm{det}_{\mathcal{T}_{1}}(\hat{O}[C])]
	\end{align}
	where, $\textrm{det}_{\mathcal{T}_1}(\hat{O}[C])$ denotes the functional determinant of the operator $\hat{O}[C]$ and, `$\mathcal{T}_1$' in the subscript is to remind us that the range of $t-$integration has been deformed to the contour $\mathcal{T}_1$ in \ref{fig:1}.

	To obtain the backreaction equation, we demand that $\delta\textrm{Re}S_{eff}/\delta C=0$, with the variation of $C$ at the endpoints assumed to be vanishing. The variation of the first part of $S_{eff}[C]$, as is given in the right-hand side of \ref{actionforCq}, is straightforward. It gives the equation of motion of $C$ when the interaction with $q$ is switched off. The variation of the second part is expected to contain the backreaction of the quantum fluctuations of $q$ on $C$. In order to find this term, we have to essentially evaluate the functional derivative of $\log[\textrm{det}_{\mathcal{T}_{1}}(\hat{O}[C])]$. It turns out that, this functional derivative can be explicitly evaluated and the final result is given by (for the full derivation, see \ref{funcderiv}):
	\begin{align}\label{deldetdelc1}
	i\frac{\delta\log[\textrm{det}_{\mathcal{T}_1}(\hat{O}[C])]}{\delta C(t)}&=\int_{-\infty e^{i\epsilon}}^{\infty}dt''\int_{-\infty e^{i\epsilon}}^{t''}dt'\left[\frac{m(C(t'))f_{in}^{*2}(t';C)}{m(C(t''))f_{in}^{*2}(t'';C)}\right]\\\nonumber
	&\times\left\{\left[\frac{\delta\omega^2(C)}{\delta C}\bigg |_{t'}+\frac{\partial_t{f_{in}^{*}}(t';C)}{f_{in}^{*}(t';C)}\frac{\delta\mu(C)}{\delta C}\bigg |_{t'}\right]\delta (t-t')\right.\\\nonumber
	&\left.-\frac{d}{dt'}\left[\frac{\partial_t{f_{in}^{*}}(t';C)}{f_{in}^{*}(t';C)}\frac{\delta\mu(C)}{\delta \dot{C}}\bigg |_{t'}\right]\delta(t-t')\right\}\\\nonumber
	&+\int_{-\infty e^{i\epsilon}}^{\infty e^{i\epsilon}}\frac{dt''}{m(C(t''))f_{in}^{*2}(t'';C)}\left[\frac{\partial_t{f_{in}^{*}}(t'';C)}{f_{in}^{*}(t'';C)}\frac{\delta\mu(C)}{\delta \dot{C}}\bigg |_{t''}\right]\delta(t-t'')
	\end{align}
	where, $f_{in}^{*}(t;C)$ is a solution of the differential equation\footnote[1]{A remark on notation: A semicolon followed by $C$ in the argument of a function (for example, $h(t;C)$) indicates functional dependence on $C$.}
	\begin{align}
	(\hat{O}[C])f_{in}^{*}(t;C)=0
	\end{align} 
	satisfying the boundary condition
	\begin{align}
	\lim_{z\rightarrow-\infty e^{i\epsilon}}f_{in}^{*}(z;C)=0.
	\end{align}
	The function $f_{in}^*(t;C)$ is nothing but the `in-mode' (i.e., positive frequency solutions at asymptotic past) of the time dependent harmonic oscillator $q$ in the background of $C$. After using the properties of $f_{in}^*(t;C)$ and a bit of algebra (see \ref{funcderiv} for details), the expression for the functional derivative can be further simplified to yield:
	\begin{align}\label{detderfinalio}
	\frac{\delta\log[\textrm{det}_{\mathcal{T}_1}(\hat{O}[C])]}{\delta C(t)}&=\partial_{C}(m^{-1})\frac{\braket{\textrm{out}|p^2(t)|\textrm{in}}}{\braket{\textrm{out}|\textrm{in}}}+\partial_C(m\omega^2)\frac{\braket{\textrm{out}|q^2(t)|\textrm{in}}}{\braket{\textrm{out}|\textrm{in}}}
	\end{align}
	where, $\ket{\textrm{in}}$ and $\ket{\textrm{out}}$ are, respectively, the `in-vacuum' and the `out-vacuum' of the $q-$subsystem interacting with the background $C(t)$.
	
	Using \ref{detderfinalio} in \ref{actionforCq}, the backreaction equation for $C$ that follows from $\delta \textrm{Re}[S_{eff}]\delta C=0$ is given by 
	\begin{align}\label{BReq}
	M\left[\ddot{C}+V'(C)\right]+\textrm{Re}\left[\frac{\partial_{C}(m^{-1})}{2}\frac{\braket{\textrm{out}|p^2(t)|\textrm{in}}}{\braket{\textrm{out}|\textrm{in}}}+\frac{\partial_C(m\omega^2)}{2}\frac{\braket{\textrm{out}|q^2(t)|\textrm{in}}}{\braket{\textrm{out}|\textrm{in}}}\right]=0
	\end{align}
	This is indeed the backreaction equation in the `in-out' approach. As alluded to before, the backreaction equation is equivalent to replacing the quantum operators acting on the Hilbert space of $q$ by a normalized `in-out' matrix element. Hence, it is non-causal owing to the presence of $\ket{\textrm{out}}$. 
	
	Another undesirable feature of this approach is that the energy conservation equation that follows from \ref{BReq} does not completely incorporate the effects of particle production. To see this, consider $\Delta E_{C}$, the total energy change of the $C$-subsystem from the asymptotic past to asymptotic future, which can be shown\cite{Rajeev:2017uwk} to be:
	\begin{align}\label{econsinout}
	\Delta E_{C}=-\frac{1}{2}\left\{\omega[C(\infty)]-\omega[C(-\infty)]\right\}
	\end{align}
	The right-hand side of \ref{econsinout} \textit{only} accounts for the change in instantaneous ground state energies of the time-dependent oscillator $q$, evaluated at times $t=-\infty$ and $t=\infty$. In quantum field theory, this manifests as the so-called vacuum-polarization effects, which may be understood as essentially being caused by the \textit{virtual} pairs produced and annihilated in the vacuum. However, in the presence of an external field, there is a non-zero probability for creation of \textit{real} particle pairs, the effects of which are expected to appear as a corresponding term in the energy conservation equation. It is clear that \ref{econsinout} does not have such a term and hence, does not incorporate the full effects of pair production. 
	
	It can be shown that these shortcomings can be remedied by simply replacing the `in-out' matrix elements in \ref{BReq} with the `in-in' expectation value, and this defines the `in-in' approach.  However, such an ad-hoc prescription seems far from rigorous. Our aim is now to give a formal basis for the `in-in' backreaction prescription through a path integral formalism. For that, we shall consider the backreaction equation, which arises when the analysis of this section is repeated for the time contour $\mathcal{T}_{2}$ in \ref{fig:2}. It is worth mentioning that the parts of $\mathcal{T}_2$, below and above the real $t-$axis, has been separately considered in the literature to represent, respectively, the forward and backward directions of time in the context of Schwinger-Keldysh formalism for a single variable (see, for instance, \cite{Kaya:2012nn}). What we aim to achieve in this work is to explicitly show that results in the `in-in' backreaction approach, for a $C-q$ system described by \ref{Cq_lag}, follows simply from the natural generalization of results in this section for the time evolution along $\mathcal{T}_2$.   
	
	\section{The `in-in' approach from complex time contour $\mathcal{T}_2$}
	We saw in the previous section that the `in-out' backreaction equation follows from the variation of the effective action $S_{eff}[C]$ that was derived by assuming that the evolution of the quantum variable $q$ was along a complex time contour $\mathcal{T}_1$. A natural question to ask at this stage is the following: Can we generalize this approach to find the effective action, say $S_{eff}^{\mathcal{T}}[C]$, for evolution along an arbitrary time contour $\mathcal{T}$ in the complex $t-$plane. The formal definition of such an effective action will be given by:
	\begin{align}\label{actionforCq2}
	S_{eff}^{\mathcal{T}}[C]=M\int_{\mathcal{T}} dt \left[\frac{\dot{C}^2}{2}-V(C)\right]+\frac{i}{2}\log[\textrm{det}_{\mathcal{T}}(\hat{O}[C])]
	\end{align}
	where the integral is along the contour $\mathcal{T}$. The effective classical evolution of $C$ along $\mathcal{T}$ can then be defined as the solution of the equation $\delta \textrm{Re}[S_{eff}^{\mathcal{T}}]/\delta C=0 $. The only non-trivial step to derive this equation is the evaluation of the functional derivative of $\log[\textrm{det}_{\mathcal{T}}(\hat{O}[C])]$. Natural generalization, of relevant standard results for evolution along real $t-$axis, to that along a complex time contour $\mathcal{T}$ allows us to show that (see \ref{funcderiv} for details):
	\begin{align}\label{deldetdelcz2}
	i\frac{\delta\log[\textrm{det}_{\mathcal{T}}(\hat{O}[C])]}{\delta C(z)}&=\int_{\mathcal{T}\vert_{z}}dz''\int_{\mathcal{T}\vert_{z''}}dz'\left[\frac{m(C(z'))f_{\sigma}^{*2}(z';C)}{m(C(z''))f_{\sigma}^{*2}(z'';C)}\right]\\\nonumber
	&\times\left\{\left[\frac{\delta\omega^2(C)}{\delta C}\bigg |_{z'}+\frac{D_z{f_{\sigma}^{*}}(z';C)}{f_{\sigma}^{*}(z';C)}\frac{\delta\mu(C)}{\delta C}\bigg |_{z'}\right]\delta (z-z')\right.\\\nonumber
	&\left.-D_{z'}\left[\frac{D_z{f_{\sigma}^{*}}(z';C)}{f_{\sigma}^{*}(z';C)}\frac{\delta\mu(C)}{\delta \dot{C}}\bigg |_{z'}\right]\delta(z-z')\right\}\\\nonumber
	&+\int_{\mathcal{T}}\frac{dz''}{m(C(z''))f_{\sigma}^{*2}(z'';C)}\left[\frac{D_z{f_{\sigma}^{*}}(z'';C)}{f_{\sigma}^{*}(z'';C)}\frac{\delta\mu(C)}{\delta \dot{C}}\bigg |_{z''}\right]\delta(z-z'')
	\end{align}
	where, $\int_{\mathcal{T}\vert_{z}}$ denotes the contour integral along $\mathcal{T}$ till the point $z$ and, $D_z$ is the directional derivative along $\mathcal{T}$. The function $f_{\sigma}$ is a solution of the differential equation $\hat{O}[C]f=0$ with the initial condition $f_{\sigma}(z_i;C)=0$, where $z_i$ is the initial point of the contour. It is easy to verify that $f_{\sigma}$ reduces to $f_{in}$ when we choose the contour to be $\mathcal{T}_1$ and we reproduce the backreaction equation for the `in-out' approach. 
	
	We shall now focus on the case when the contour is chosen to be $\mathcal{T}_2$ in \ref{fig:2}. Since, $\mathcal{T}_1$ and $\mathcal{T}_2$ coincide asymptotically in the beginning, it turns out that $f_{\sigma}$ is precisely $f_{in}$ for the choice $\mathcal{T}=\mathcal{T}_2$ as well. 
	This implies that the generalization of \ref{deldetdelc1}, to the case where time evolution is along the complex contour $\mathcal{T}_2$, is given by: 
	\begin{align}\label{deldetdelcz}
	i\frac{\delta\log[\textrm{det}_{\mathcal{T}_2}(\hat{O}[C])]}{\delta C(z)}&=\int_{\mathcal{T}_2\vert_{z}}dz''\int_{\mathcal{T}_2\vert_{z''}}dz'\left[\frac{m(C(z'))f_{in}^{*2}(z';C)}{m(C(z''))f_{in}^{*2}(z'';C)}\right]\\\nonumber
	&\times\left\{\left[\frac{\delta\omega^2(C)}{\delta C}\bigg |_{z'}+\frac{D_z{f_{in}^{*}}(z';C)}{f_{in}^{*}(z';C)}\frac{\delta\mu(C)}{\delta C}\bigg |_{z'}\right]\delta (z-z')\right.\\\nonumber
	&\left.-D_{z'}\left[\frac{D_z{f_{in}^{*}}(z';C)}{f_{in}^{*}(z';C)}\frac{\delta\mu(C)}{\delta \dot{C}}\bigg |_{z'}\right]\delta(z-z')\right\}\\\nonumber
	&+\int_{\mathcal{T}_2}\frac{dz''}{m(C(z''))f_{in}^{*2}(z'';C)}\left[\frac{D_z{f_{in}^{*}}(z'';C)}{f_{in}^{*}(z'';C)}\frac{\delta\mu(C)}{\delta \dot{C}}\bigg |_{z''}\right]\delta(z-z'')
	\end{align}
	where, $\int_{\mathcal{T}_2\vert_{z}}$ denotes a contour integral along $\mathcal{T}_2$ till the point $z$ and $D_{z}$ denotes the directional derivative along $\mathcal{T}_2$. Once again, we have delegated the details to \ref{funcderiv}. 
	
	This choice of contour $\mathcal{T}_2$ indeed gives us the `in-in' backreaction equation. The expression for functional derivative in \ref{deldetdelcz} can be further simplified to give (see \ref{funcderiv} for details)
	\begin{align}\label{deldetdel4}
	i\frac{\delta\log[\det_{\mathcal{T}_{2}}(\hat{O}[C])]}{\delta C(z)}&=-\left[\left(\partial_{C}m\right)\dot{f}_{in}(z;C)D_z{f}^*_{in}(z;C)-\partial_C(m\omega^2)f_{in}(z;C)f^*_{in}(z;C)\right]
	\end{align}
	From \ref{fig:2}, we can see that a point, say $t$, in the real time axis gets mapped to two points on $\mathcal{T}_2$, say $t_{-}$ and $t_{+}$, which we can identify as the forward and backward evolution in time, respectively. Further, the doublet $\{C(t_{+}),C(t_-)\}$ which can be constructed out of variable $C(z)$ for $z\in \mathcal{T}_2$, is reminiscent of the `doubled degrees of freedom' $\{C_{+}(t), C_{-}(t)\}$ akin to the Schwinger-Keldysh formalism, but it arises rather naturally in our approach. Thus, the effects of this `doubling' are implicitly incorporated in our approach by virtue of the specific form of $\mathcal{T}_2$.  In the conventional Schwinger-Keldysh approach, in the classical limit, the equation of motion of $C$ is retained by making the identification $C_{+}(t)=C_{-}(t)=C(t)$, \textit{after} the variational principle is applied. Along similar lines, the backreaction equation that governs the effective classical dynamics of $C$, in our approach, can be obtained by demanding $\lim_{\epsilon\rightarrow 0}C(t_{+})=\lim_{\epsilon\rightarrow 0}C(t_{-})=C(t)$ in \ref{deldetdel4}. This procedure, along with the results
	\begin{align}
	\braket{\textrm{in}|q^2(t)|\textrm{in}}&=f_{in}(t;C)f_{in}^{*}(t;C);&&\braket{\textrm{in}|p^2(t)|\textrm{in}}=m^2\dot{f}_{in}(t;C)\dot{f}_{in}^*(t;C),
	\end{align}
	finally yields the following form for the backreaction equation:
	\begin{align}\label{BReqinin}
	M\left[\ddot{C}+V'(C)\right]+\frac{\partial_{C}(m^{-1})}{2}\braket{\textrm{in}|p^2(t)|\textrm{in}}+\frac{\partial_C(m\omega^2)}{2}\braket{\textrm{in}|q^2(t)|\textrm{in}}=0
	\end{align}
	Therefore, we retain the `in-in' backreaction equation as claimed. It is worth mentioning that this equation is causal. 
	
	Multiplying both sides of \ref{BReqinin} by $\dot{C}$ and simplifying the equation we get the energy conservation law: 
	\begin{align}
	\frac{d}{dt}\left[M\frac{\dot{C}^2}{2}+M V(C)+\left(\frac{1}{2m}\braket{\textrm{in}|p^2(t)|\textrm{in}}+\frac{m\omega^2}{2} \braket{\textrm{out}|q^2(t)|\textrm{in}}\right)\right]=0
	\end{align}
	This conservation equation was also discussed in \cite{Rajeev:2017uwk}. It was shown that energy conservation equation can be written in terms of the mean number $n(t)$ of particles produced as:
	\begin{align}
	\label{quadratice_energy_inin}
	\frac{d}{dt}\left[\frac{M}{2}\dot{C}^{2}+MV(C)+\left(n+\frac{1}{2}\right)\omega(C)\right]=0
	\end{align}
	This equation can be intuitively understood as follows: the backreaction on $C$ from the quantum degree of freedom has two parts: (i) one coming from the particle production of $q$ system, namely $d(n\omega)/dt$ and (ii) the other coming from the change in vacuum energy of $q$ due to the interaction with $C$, namely $d(\omega/2)/dt$. Note that, in sharp contrast with the energy conservation that followed from the `in-out' prescription, which did not take into account the effects of particle production.
	
	\section{Discussion}
	The backreaction of a quantum degree of freedom on an effectively classical system is ubiquitous in physics; it is relevant in the study of black hole evaporation by Hawking radiation and structure formation in the early universe, just to name a few. For a system composed of an effectively classical part ($C$) coupled to a quantum degree of freedom ($q$), a straightforward application of the semi-classical analysis using path integral formalism gives the so-called `in-out' backreaction equation. This approach has two serious pathologies, viz., (i) non-causal evolution and (ii) an unphysical energy-conservation equation. A natural alternative is the so-called `in-in' approach, which is devoid of these shortcomings of the `in-out' approach. Our main goal in this work was to derive `in-in' backreaction directly from path integral formalism.
	
	We considered a specific $C-q$ system in this work, in which the quantum part $q$ is essentially a time-dependent harmonic oscillator, for a fixed background configuration $C(t)$ of the classical subsystem $C$. When the evolution is along the $\mathcal{T}_1$ of \ref{fig:1}, we show that the corresponding back reaction equation, obtained by varying the effective action $S^{\mathcal{T}_1}_{eff}[C]$, matches exactly with that of the `in-out' formalism. On the other hand, for the choice of time-contour $\mathcal{T}_2$ of \ref{fig:2}, the backreaction obtained by varying the corresponding effective action $S^{\mathcal{T}_2}_{eff}[C]$ turns out to be precisely that of the `in-in' formalism. Therefore, we have provided a path integral based approach for deriving the correct backreaction prescription which: (i) is causal and (ii) has the correct form of energy conservation equation. 
	
	Our approach based on the concept of evolution along complex time contours also provides a unified formalism for studying both the `in-out' and `in-in' backreaction equation. The effective classical equation of motion for evolution of $C$ along a complex time contour $\mathcal{T}$ can be written as:
	\begin{align}\label{breq}
	\frac{\delta\textrm{Re}[S_{eff}^{\mathcal{T}}[C]]}{\delta C}=0
	\end{align}
	where, the effective action $S_{eff}^{\mathcal{T}}[C]$ is formally defined by \ref{actionforCq2}. From this \textit{single} general equation, `in-in' and `in-out' approaches can be derived in a unified manner. When we chose the complex time contour to be $\mathcal{T}_1$, the equation of motion \ref{breq} implies
	\begin{align}
	M\ddot{C}+MV'(C)+\textrm{Re}\left[\frac{\partial_{C}(m^{-1})}{2}\frac{\braket{\textrm{out}|p^2(t)|\textrm{in}}}{\braket{\textrm{out}|\textrm{in}}}+\frac{\partial_C(m\omega^2)}{2}\frac{\braket{\textrm{out}|q^2(t)|\textrm{in}}}{\braket{\textrm{out}|\textrm{in}}}\right]=0,
	\end{align}
	On the other hand, when we chose the complex time contour be $\mathcal{T}_2$, the equation of motion \ref{breq} gives
	\begin{align}
	M\ddot{C}+MV'(C)+\frac{\partial_{C}(m^{-1})}{2}\braket{\textrm{in}|p^2(t)|\textrm{in}}+\frac{\partial_C(m\omega^2)}{2}\braket{\textrm{in}|q^2(t)|\textrm{in}}=0.
	\end{align}

	\section*{Acknowledgement}
	I thank Prof. T. Padmanabhan for discussions and detailed comments on the draft. I am indebted to the two anonymous referees for their critical reading of the manuscript and valuable comments which have greatly helped in improving the presentation of the results in this work. My research is supported by
	Senior Research Fellowship of the Council of Scientific and Industrial Research (CSIR), India.
	
	\appendix
	\labelformat{chapter}{Appendix #1}
	\labelformat{section}{Appendix #1}
	\labelformat{subsection}{Appendix #1}
	\labelformat{subsubsection}{Appendix #1}
	\section*{Appendix}
	\label{AppendixG} % Change X to a consecutive letter; for referencing this appendix elsewhere, use \ref{AppendixX}
	The derivations of results that are directly relevant to the main body of this paper are given in \ref{funcderiv} and \ref{inoutcorrelator}. In \ref{GY}, we review some of the standard results concerning the quantum mechanics of a TDHO. In \ref{redorder}, we briefly discuss some mathematical results related to solutions of a TDHO equation that are of useful in some of the derivations in \ref{funcderiv} and \ref{inoutcorrelator}. 
	\section{Derivation of the Gel'fand-Yaglom formula}\label{GY}
	Consider a TDHO with the following Lagrangian:
	\begin{align}
	\mathcal{L}_{q}=\frac{m(t)}{2}\left(\dot{q}^2-\omega^2(t)q^2\right).
	\end{align}
	The classical dynamics of the TDHO is governed by the equation of motion corresponding to this Lagrangian, which is given by
	\begin{align}\label{classeom}
	\hat{O}q \equiv \left[\frac{d}{dt}\left(m(t)\frac{d}{dt}\right)+m(t)\omega^2(t)\right]q=0.
	\end{align}
	The quantum dynamics, on the other hand, is encoded in the Schr\"{o}dinger propagator $\mathcal{G}_{q}(q_f,t_f;q_i,t_i)$ for this system, which takes the form \cite{chaichian2018path,feynman1979path}:
	\begin{align}\label{kernelqm}
	\mathcal{G}_{q}(q_f,t_f;q_i,t_i)= e^{i S_{cl}(q_f,t_f;q_i,t_i)}\underbrace{\int \mathcal{D}y \exp\left(-\frac{i}{2}\int_{t_i}^{t_f}dt \,y(t)\hat{O}y(t)\right)}_{\equiv \mathcal{F}_{q}(t_f,t_i)}
	\end{align} 
	where, $S_{cl}$ is the action evaluated at the classical path (i.e., the solution of \ref{classeom}) that starts from $q_i$ at $t=t_i$ and ends on $q_f$ at $t=t_f$ and, the paths $y:I\equiv[t_i,t_f]\rightarrow \mathbb{R}$ satisfy the following boundary conditions:
	\begin{align}\label{boundarycond}
	y(t_i)=y(t_f)=0.
	\end{align}
	The path integral over all $y$ in \ref{kernelqm}, which we have denoted by $\mathcal{F}_{q}$, may also be formally written as
	\begin{align}\label{defandF}
	\mathcal{F}_{q}(t_f,t_i)\propto \left[\textrm{det}_{\mathrm{I}}(\hat{O})\right]^{-1/2}
	\end{align}
	where, $\textrm{det}_{\mathrm{I}}()$ denotes the determinant of the projection of an operator in the subspace of all functions $y:I\rightarrow \mathbb{R}$, that satisfy the boundary condition in \ref{boundarycond}. Such a determinant, in general, is a divergent quantity. However, the following ratio is finite and well defined:
	\begin{align}\label{detandG}
	\frac{\det_{\mathrm{I}}(\hat{O}_2)}{\det_{\mathrm{I}}(\hat{O}_1)}=\left[\frac{\mathcal{F}_{q_{1}}(t_f,t_i)}{\mathcal{F}_{q_2}(t_f,t_i)}\right]^2=\left[\frac{\mathcal{G}_{q_1}(0,t_f;0,t_i)}{\mathcal{G}_{q_2}(0,t_f;0,t_i)}\right]^2
	\end{align} 
	where,
	\begin{align}    \hat{O}_1=\frac{d}{dt}\left(m_1(t)\frac{d}{dt}\right)+m_1(t)\omega_1^2(t);\qquad \textrm{and}&&
	\hat{O}_2=\frac{d}{dt}\left(m_2(t)\frac{d}{dt}\right)+m_2(t)\omega_2^2(t),
	\end{align}
	and, $\mathcal{G}_{q_i}$ is the Schr\"{o}dinger propagator for a harmonic oscillator of mass $m_i(t)$ and frequency $\omega_i(t)$. Using \ref{detandG}, one can calculate the ratio of determinants of two operators $\hat{O}_1$ and $\hat{O}_2$ from the respective Schr\"{o}dinger kernels, $\mathcal{G}_{q_1}$ and $\mathcal{G}_{q_2}$. Much of our analysis in the later sections require the explicit form of $\textrm{det}_{\mathrm{I}}(\hat{O})$ for the most general TDHO, which in tern requires the explicit form of $\mathcal{G}_{q}$. Several standard references (for instance, \cite{chaichian2018path}) on path integral formulation of quantum mechanics do provide useful expressions for $\textrm{det}_{\mathrm{I}}(\hat{O})$. However, for completeness, we present a simpler derivation here.

	In \cite{Padmanabhan:2017rsj} (also see \cite{RoblesPerez:2017gsc}), it was shown that a TDHO described by \ref{classeom} can be mapped to a simple harmonic oscillator (SHO), $Q$, of unit mass and constant frequency $\Omega$. This is achieved by first defining $Q=q/f$, where $f$ satisfies
	\begin{align}\label{ermokov2}
	m f^3\left[\frac{d}{dt}(m\dot{f})+m f^2\right]=\Omega^2
	\end{align}
	Then it can be shown that, in terms of a new time coordinate $\tau$ defined by $d\tau=dt/(mf^2)$, the equation of motion of $Q$ reduces to
	\begin{align}
	\frac{d^2Q}{d\tau^2}+\Omega^2Q^2=0
	\end{align}
	This mapping, in the quantum mechanical description, translates to the following relation between the Schr\"{o}dinger propagators of the two systems:
	\begin{align}
	&\mathcal{G}_{q}(q_f,t_f;q_i,t_i)=\\\nonumber
	&\frac{1}{\sqrt{f(t_f)f(t_i)}}\exp\left[im(t_f)\frac{\dot{f}(t_f)}{f(t_f)}q_f^2-im(t_i)\frac{\dot{f}(t_i)}{f(t_i)}q_i^2\right]
	\mathcal{G}_{Q}\left[\frac{q_f}{f(t_f)},\tau(t_f); \frac{q_i}{f(t_i)},\tau(t_i)\right].
	\end{align}
	But, the propagator $\mathcal{G}_{Q}$ for a SHO of unit mass and constant frequency $\Omega$ is well known and is given by
	\begin{align}
	&\mathcal{G}_{Q}(Q_f,\tau_f;Q_1,\tau_i)=\\\nonumber
	&\left(\frac{\Omega}{2\pi i \sin[\Omega(\tau_f-\tau_i)]}\right)^{1/2}\exp \left(-{\frac {\Omega \left\{(Q_f^{2}+Q_i^{2})\cos[\Omega(\tau_f-\tau_i)]-2Q_fQ_i\right\}}{2i\sin[\Omega(\tau_f-\tau_i)]}}\right).
	\end{align}
	Therefore, we arrive at the following convenient expression for $\mathcal{G}_{q}$:
	\begin{align}\label{proptdho}
	\mathcal{G}_{q}(q_f,t_f;q_i,t_i)=\left(\frac{2\pi i f(t_f)f(t_i) \sin\left[\Omega\int_{t_i}^{t_f}\frac{dt'}{m(t')f(t')^2}\right]}{\Omega}\right)^{-1/2} e^{i S_{cl}(q_f,t_f;q_i,t_i)}
	\end{align}
	This equations gives the propagator for a general TDHO in terms of a single function $f$. Therefore, once we solve \ref{ermokov2} for $f$, \ref{proptdho} may be used to obtain the explicit form of the propagator $\mathcal{G}_{q}$, from which one can obtain $\mathcal{F}_{q}$ or equivalently $\det_{\mathrm{I}}(\hat{O})$. 
	
	However, it turns out that one does not necessarily need the full expression for $\mathcal{G}_{q}$ in order to obtain $\mathcal{F}_{q}$. We will now show that $\mathcal{F}_{q}^{-2}\propto \det_{\mathrm{I}}(\hat{O})$ can be obtained from a particular solution of a rather simple differential equation, without ever deriving the explicit form of the propagator. 
	
	In the light of \ref{proptdho}, it is convenient to define the following quantity: 
	\begin{align}
	g(t;t_i)\equiv\frac{f(t)f(t_i) \sin\left[\Omega\int_{t_i}^{t}\frac{dt'}{m(t')f(t')^2}\right]}{\Omega}
	\end{align}
	The path integral term $\mathcal{F}_{q}$ in the propagator is related to this quantity via $g(t_f;t_i)=\mathcal{F}_{q}^{-2}\propto \det_{\mathrm{I}}(\hat{O})$.
	It can be verified using \ref{ermokov2} that $g(t;t_i)$ is a solution of the TDHO equation given by \ref{classeom}, with the following initial conditions: (i) $    q(t_i)=0$ and (ii) $\dot{q}(t_i)=[m(t_i)]^{-1}$.
	Once, this particular solution is obtained, we can immediately obtain $g(t_f;t_i)$ or equivalently $\mathcal{F}_{q}$. To summarize, we have the following result:
	\begin{align}\label{detandg}
	\frac{\det_{\mathrm{I}}(\hat{O}_2)}{\det_{\mathrm{I}}(\hat{O}_1)}=\frac{g_2(t_f;t_i)}{g_1(t_f;t_i)}
	\end{align}
	where $g_{j}(t;t_i)$, for $j\in\{1,2\}$, is the solution of $\hat{O}_{j}g_j=0$, with the initial condition
	\begin{align}\label{initialcond}
	g_j(t_i;t_i)=0&&\textrm{and}&&\dot{g}_j(t_i;t_i)=\frac{1}{m_j(t_i)}.
	\end{align}
	The second condition in \ref{initialcond}, however, is just fixing the normalization of $g_j$. One can easily generalize \ref{detandg} for an arbitrary normalization of $g_j$ as follows.
	\begin{align}\label{detandggen}
	\frac{\det_{\mathrm{I}}(\hat{O}_2)}{\det_{\mathrm{I}}(\hat{O}_1)}=\frac{g_2(t_f;t_i)\dot{g}_1(t_i;t_i)m_1(t_i)}{g_1(t_f;t_i)\dot{g}_2(t_i;t_i)m_2(t_i)}
	\end{align}
	This is the well known Gel'fand-Yaglom formula \cite{gelfand} (also see \cite{Dowker:2011np} for a recent discussion). What we have shown here is that, using the mapping given in \cite{Padmanabhan:2017rsj}, we can derive \ref{detandg} from just the knowledge of $\mathcal{G}_{Q}$ for a SHO of unit mass and constant frequency $\Omega$. 
	
	\subsection{An important special case}
	We will see in the subsequent sections that, for the derivation of backreaction equations, we need an extension of \ref{detandg} to cases when the interval $I=[t_i,t_f]$ is infinite. If we proceed with some care, such an extension can be obtained from the results we have already discussed so far.
	
	Note that, since the operator $\hat{O}$ is Hermitian, strictly speaking, the path integral in \ref{kernelqm} is not convergent. We can circumvent this issue by adding an infinitesimal negative imaginary part to the operator, i.e., $\hat{O}\rightarrow\hat{O}-i\epsilon$, where $\epsilon\rightarrow 0^{+}$. Hence, \ref{detandg} is to be interpreted as the $\epsilon\rightarrow 0^{+}$ limit of
	\begin{align}\label{detandgepsilon}
	\frac{\det_{\mathrm{I}}(\hat{O}_2-i\epsilon)}{\det_{\mathrm{I}}(\hat{O}_1-i\epsilon)}=\frac{g^{(\epsilon)}_2(t_f;t_i)}{g^{(\epsilon)}_1(t_f;t_i)}
	\end{align}
	where $g^{(\epsilon)}_{j}(t;t_i)$, for $j=1,2$, is the solution of $(\hat{O}_{j}-i\epsilon)g^{(\epsilon)}_j=0$ with the initial conditions: (i) $g^{(\epsilon)}_j(t_i;t_i)=0$ and (ii) $\dot{g}^{(\epsilon)}_j(t_i;t_i)=[m_j(t_i)]^{-1}$.
	When the interval $I$ is finite, the limiting procedure is trivial. In fact, in this case, we can use \ref{detandg} directly without any harm. However, when $I=(-\infty,\infty)=\mathbb{R}$, the `$i\epsilon-$prescription' needs to be employed carefully. 
	
	To this end, let us first consider two TDHOs, namely $q_1$ and $q_2$, with time dependent masses and frequencies given by
	\begin{align}\label{defmoft}
	m_{j}(t)=\begin{cases}
	m_j(\eta_i)=m_0;\quad &(\textrm{for }t_i<t<\eta_i)\\
	m_{j}(t);\quad&(\textrm{for }\eta_i<t<\eta_f)\\
	m_{j}(\eta_f)=M_0;\quad&(\textrm{for }\eta_f<t<t_f)
	\end{cases}
	\end{align}  
	and
	\begin{align}\label{defomegaoft}
	\omega_{j}(t)=\begin{cases}
	\omega_j(\eta_i)=\omega_0;\quad&(\textrm{for }t_i<t<\eta_i)\\
	\omega_{j}(t);\quad&(\textrm{for }\eta_i<t<\eta_f)\\
	\omega_{j}(\eta_f)=\Omega_0;\quad&(\textrm{for }\eta_f<t<t_f)
	\end{cases}
	\end{align} 
	for $j\in\{1,2\}$. That is, the frequencies and masses of both the oscillators have the \textit{same} constant asymptotic values for $t$ outside the interval $\tilde{I}=(\eta_i,\eta_f)$. This restricted class of $m_{i}(t)$ and $\omega_{j}(t)$, as we will see shortly, turns out to be exactly the class of time dependent mass and frequency that is required for the derivation of backreaction equations. We now seek for the positive frequency solutions $f^{(\epsilon)}_{j(in)}(t)$ in the past  (i.e., the ones that behave as $\sim e^{-i\omega_0 t}$ near $t=t_i$), of the set of TDHO equations 
	\begin{align}\label{iepsiloneom}
	(\hat{O}_{j}-i\epsilon)q_j=0.
	\end{align}
	The exact form of these solutions can be written in terms of time-dependent Bogoluibov coefficients as follows: First, for $t_i<t<\eta_i$ we have
	\begin{align}\label{finline1}
	f^{(\epsilon)}_{j(in)}(t)=\dfrac{e^{-i\left(\omega_0-i\frac{\epsilon}{2\omega_0^2m_0^2}\right)(t-\eta_i)}}{\sqrt{m_0\omega_0}};\quad(\textrm{for }t_i<t<\eta_i).
	\end{align}
	Second, $\textrm{for }\eta_i<t<\eta_f$ we have
	\begin{align}\label{finline2}
	f^{(\epsilon)}_{j(in)}(t)=\dfrac{\alpha_{j}(t)}{\sqrt{m_{j}(t)\omega_j(t)}}e^{-i\rho_j}+\dfrac{\beta_{j}(t)}{\sqrt{m_{j}(t)\omega_j(t)}}e^{i\rho_j};\quad(\textrm{for }\eta_i<t<\eta_f)
	\end{align}
	Finally, $\textrm{for }\eta_f<t<t_f$ we have
	\begin{align}\label{finline3}
	f^{(\epsilon)}_{j(in)}(t)&=\left\{\dfrac{\alpha_{j}(\eta_f)}{\sqrt{M_0\Omega_0}}e^{-i\rho_j(\eta_f)}\right\}e^{-i\left(\Omega_0-i\frac{\epsilon}{2\Omega_0^2M_0^2}\right)(t-\eta_f)}\\\nonumber
	&+\left\{\dfrac{\beta_{j}(\eta_f)}{\sqrt{M_0\Omega_0}}e^{i\rho_j(\eta_f)}\right\}e^{i\left(\Omega_0-i\frac{\epsilon}{2\Omega_0^2M_0^2}\right)(t-\eta_f)};\quad(\textrm{for }\eta_f<t<t_f)
	\end{align}
	where, $\dot{\rho}_j=(\omega_j-i\epsilon/(2m_j^2\omega_j^2))$ and, $\alpha_{j}(t)$ and $\beta_{j}(t)$ are time dependent Bogoluibov coefficients\cite{Mahajan:2007qc} that satisfy
	\begin{align}
	\dot{\alpha}_{j}&=\frac{1}{2}\left(\frac{\dot{\omega}_{j}}{\omega_{j}}+\frac{\dot{m}_{j}}{m_{j}}\right)\beta_{j}e^{2i\rho_{j}}\\
	\dot{\beta}_{j}&=\frac{1}{2}\left(\frac{\dot{\omega}_{j}}{\omega_{j}}+\frac{\dot{m}_{j}}{m_{j}}\right)\alpha_{j}e^{-2i\rho_{j}}
	\end{align}
	\ref{finline3} shows that positive frequency solutions in the past evolve into linear combination of positive and negative frequency solutions in the future. Now, the negative frequency solutions in the past $\tilde{f}_j$ can be similarly found as follows: First $\textrm{for }(t_i<t<\eta_i)$ we have
	\begin{align}\label{ftilde1}
	\tilde{f}^{(\epsilon)}_{j(in)}(t)=\dfrac{e^{i\left(\omega_0-i\frac{\epsilon}{2\omega_0^2m_0^2}\right)(t-\eta_i)}}{\sqrt{m_0\omega_0}};\quad(\textrm{for }t_i<t<\eta_i)
	\end{align}
	Second, $\textrm{for }\eta_i<t<\eta_f$ we get
	\begin{align}\label{ftilde2}
	\tilde{f}^{(\epsilon)}_{j(in)}(t)=\dfrac{\alpha^*_{j}(t)}{\sqrt{m_{j}(t)\omega_j(t)}}e^{i\rho_j}+\dfrac{\beta^*_{j}(t)}{\sqrt{m_{j}(t)\omega_j(t)}}e^{-i\rho_j};\quad(\textrm{for }\eta_i<t<\eta_f)
	\end{align}
	Finally, $\textrm{for }\eta_f<t<t_f$ we have
	\begin{align}\label{ftilde3}
	\tilde{f}^{(\epsilon)}_{j(in)}(t)&=\left\{\dfrac{\alpha^*_{j}(\eta_f)}{\sqrt{M_0\Omega_0}}e^{i\rho_j(\eta_f)}\right\}e^{i\left(\Omega_0-i\frac{\epsilon}{2\Omega_0^2M_0^2}\right)(t-\eta_f)}\\\nonumber
	&+\left\{\dfrac{\beta_{j}^*(\eta_f)}{\sqrt{M_0\Omega_0}}e^{-i\rho_j(\eta_f)}\right\}e^{-i\left(\Omega_0-i\frac{\epsilon}{2\Omega_0^2M_0^2}\right)(t-\eta_f)};\quad(\textrm{for }\eta_f<t<t_f)
	\end{align}
	The index $j$ (taking values $1$ and $2$) in the subscript denotes which oscillator, among $q_1$ and $q_2$, we are referring to. As regards the subscript `$(in)$', first recall that `in-modes' refer to the solutions of a TDHO that behave as positive frequency modes in the asymptotic past and `out-modes' are those that behave as positive frequency modes in the asymptotic future. With this definition in mind, it is easy to see that: (i) the solutions $f^{(\epsilon)}_{j(in)}(t)$, in the $\epsilon\rightarrow 0$ limit, are just the `in-modes' (hence, the subscript) and (ii) the `tilde' symbol, in the $\epsilon\rightarrow 0$ limit, represent complex-conjugation. Now, given these functions, we can make the following claim.\\
	\vspace{.2cm}\\
	\textbf{Claim:} Let $\hat{O}_{1}$ and $\hat{O}_{2}$ be the equation of motion operator for two TDHOs $q_1$ and $q_2$, respectively, with the time dependent mass and frequency of the form given in \ref{defmoft} and \ref{defomegaoft}, then 
	\begin{align}\label{claim}
	\lim_{(t_i,t_f)\rightarrow(-\infty,\infty)}\left[\frac{\int \mathcal{D}y\,\, e^{-\frac{i}{2}\int_{t_i}^{t_f}dt \,y\left(\hat{O}_1-i\epsilon\right)y}}{\int \mathcal{D}y\,\, e^{-\frac{i}{2}\int_{t_i}^{t_f}dt \,y\left(\hat{O}_2-i\epsilon\right)y}}\right]^2=\lim_{I\sim \mathbb{R}}\frac{\det_{\mathrm{I}}(\hat{O}_2-i\epsilon)}{\det_{\mathrm{I}}(\hat{O}_1-i\epsilon)}=\frac{\alpha^*_2(\eta_f)e^{i\rho_2(\eta_f)}}{\alpha^*_1(\eta_f)e^{i\rho_1(\eta_f)}}
	\end{align}
	where, by $I\sim\mathbb{R}$ we mean $t_{i}\rightarrow-\infty$ and $t_{f}\rightarrow\infty$.\\
	\vspace{.2cm}\\
	\textbf{Proof:} From \ref{detandgepsilon}, the ratio of determinants of $(\hat{O}_2-i\epsilon)$ and $(\hat{O}_1-i\epsilon)$ can be found once the appropriate functions $g^{(\epsilon)}_j(t;t_i)$ are known. Note that $\tilde{f}^{(\epsilon)}_{j(in)}(t)$, by definition satisfies the differential equation $(\hat{O}_2-i\epsilon)\tilde{f}^{(\epsilon)}_{j}=0$
	and from \ref{ftilde1} we also see that $\lim_{t\rightarrow-\infty}\tilde{f}^{(\epsilon)}_{j(in)}(t)=0$.
	Hence, we identify $g^{(\epsilon)}_j(t;-\infty)=\tilde{f}^{(\epsilon)}_{j(in)}(t)$. This implies
	\begin{align}\label{detandf}
	\lim_{I\sim \mathbb{R}}\frac{\det_{\mathrm{I}}(\hat{O}_2-i\epsilon)}{\det_{\mathrm{I}}(\hat{O}_1-i\epsilon)}=\lim_{t\rightarrow\infty}\frac{\tilde{f}^{(\epsilon)}_{2(in)}(t)}{\tilde{f}^{(\epsilon)}_{1(in)}(t)}
	\end{align}
	Note that the extra normalization factors discussed in \ref{detandggen} turns out to be unity in this case owing to the identical asymptotic behaviour of the oscillators. Now, the right hand side of \ref{detandf} can be easily evaluated using \ref{ftilde3} to get
	\begin{align}\label{rationofalphas}
	\lim_{t\rightarrow\infty}\frac{\tilde{f}^{(\epsilon)}_{2(in)}(t)}{\tilde{f}^{(\epsilon)}_{1(in)}(t)}=\frac{\alpha^*_2(\eta_f)e^{i\rho_2(\eta_f)}}{\alpha^*_1(\eta_f)e^{i\rho_1(\eta_f)}}
	\end{align}
	This proves the claim.
	
	It is now worth emphasizing the importance of \ref{claim}. The left-hand side of the first equality in \ref{claim} is defined entirely in terms of path integrals that are connected with the \textit{quantum} evolution of two TDHOs $q_1$ and $q_2$. On the other hand, the right-hand side of the second equality in \ref{claim} is obtained by just solving the \textit{classical} equation of motion of the same oscillator systems. Therefore, \ref{claim} is a remarkable equation that relates the path integral formalism and the standard approach for studying the quantum evolution of TDHOs using Bogoluibov coefficients.       
	
	Another interesting interpretation for the right hand side of \ref{rationofalphas} can be obtained by noting that,
	\begin{align}\label{detintermsoffin}
	\lim_{t\rightarrow\infty}\frac{f^{*}_{2(in)}(e^{i\epsilon}t)}{f^{*}_{1(in)}(e^{i\epsilon}t)}=\frac{\alpha^*_2(\eta_f)e^{i\rho_2(\eta_f)}}{\alpha^*_1(\eta_f)e^{i\rho_1(\eta_f)}}
	\end{align}  
	where, $f^{*}_{j(in)}(t)\equiv\tilde{f}^{(\epsilon\rightarrow 0)}_{j(in)}(t)$. This seems to indicate that the `$i\epsilon-$prescription' can also be effected by rotating the time axis by a positive angle $\epsilon$ in the complex $t-$plane. Hence, \ref{claim} in this interpretation translates to:
	\begin{align}\label{claim2}
	\left[\frac{\int \mathcal{D}y\,\, e^{-\frac{i}{2}\int_{\mathcal{T}_1}dt \,y\hat{O}_1y}}{\int \mathcal{D}y\,\, e^{-\frac{i}{2}\int_{\mathcal{T}_1}dt \,y\hat{O}_2y}}\right]^2=\frac{\det_{\mathcal{T}_1}(\hat{O}_2)}{\det_{\mathcal{T}_1}(\hat{O}_1)}=\frac{\alpha^*_2(\eta_f)e^{i\rho_2(\eta_f)}}{\alpha^*_1(\eta_f)e^{i\rho_1(\eta_f)}}
	\end{align}
	where, the complex contour $\mathcal{T}_1$ is as shown in \ref{fig:1}, with $t_0=0$. This result indicates that complex time contours can be a useful tool in the study of TDHOs. We shall now explore the relevant mathematical results concerning this subject. 
	\subsection{Complex time contours}\label{PIcontour}
	We define a time contour as a continuous map from an interval in real line to the complex time plane. For example, $\mathcal{T}:(-1,1]\rightarrow \mathbb{C}$ given by
	\begin{align}
	\mathcal{T}(\tau)=e^{i\pi \tau}
	\end{align}
	is a `closed time contour'. Two special contours will be of particular interest to us: (i) $\mathcal{T}_1: \mathbb{R}\rightarrow \mathbb{C}$ and (ii) $\mathcal{T}_2: \mathbb{R}\rightarrow \mathbb{C}$, defined by (see \ref{figure})
	\begin{align}
	\mathcal{T}_1(t)&= (t-t_0)e^{i\epsilon};\,\,\,t\in\mathbb{R}\\
	\mathcal{T}_2(t)&=\begin{cases}
	(t-t_0)e^{i \epsilon};\quad&(\textrm{for }-\infty<t<t_0)\\
	(t-t_0)e^{-i\epsilon};\quad&(\textrm{for }t_0<t<\infty)
	\end{cases}
	\end{align}
	A technical comment is in order; $\mathcal{T}_2$ is not a smooth contour according to our definition, however we can treat it as the limit of an appropriate smooth curve. Having defined the basis notions of complex time contours, let us now look at the dynamics of a TDHO along them. Consider, again, the following TDHO equation:
	\begin{align}\label{difftdho}
	\frac{d}{dt}\left(m\dot{q}\right)+\omega^2(t)q=0.
	\end{align} 
	We would now like to replace the derivatives in the above differential equation with the `directional derivatives' along contours. That is, we want to replace:
	\begin{align}\label{prescription}
	\frac{d}{dt}\rightarrow\frac{1}{\dot{\mathcal{T}}(\tau)}\frac{d}{d\tau}
	\end{align}
	where, $\tau$ is a parameter along the contour $\mathcal{T}$. The differential equation hence obtained, takes the form:
	\begin{align}\label{diffeqt}
	\frac{1}{\dot{\mathcal{T}}(\tau)}\frac{d}{d\tau}\left(\frac{m(\mathcal{T}(\tau))}{\dot{\mathcal{T}}(\tau)}\frac{d q}{d\tau}\right)+\omega^2(\mathcal{T}(\tau))q=0.
	\end{align}
	where, dot denotes derivative w.r.t $\tau$ and we have assumed $\omega^2(t)$ and $m(t)$ can be analytically continued to the complex plane. Further, assuming that $q$ can be analytically continued to an open domain $\mathbb{D}\supset\mathcal{T}$, we can use the properties of an analytic function to rewrite \ref{diffeqt} as
	\begin{align}\label{diffeqwithz}
	\frac{d}{dz}\left(m(z)\frac{dq}{dz}\right)+\omega^2(z)q=0\,\,\,;\quad(\textrm{for }z\in\mathcal{T})
	\end{align}
	This implies that the analytic continuation of $q(t)$, namely $q(z)$, is a solution of the differential equation
	\begin{align}\label{diffeqwithzinD}
	\frac{d}{dz}\left(m(z)\frac{dq}{dz}\right)+\omega^2(z)q=0\,\,\,;\quad(\textrm{for }z\in\mathbb{D})
	\end{align}
	Hence, we immediately obtain the following simple result:
	\\
	\vspace{.3cm}\\
	\textbf{Result:} If $h$ is a solution of \ref{diffeqwithzinD} then $h(\mathcal{T}(\tau))$ is a solution of \ref{diffeqt}.
	\\
	\vspace{.3cm}
	
	Let us now look at the quantum evolution of the TDHO along an arbitrary contour $\mathcal{T}$. For $\mathcal{T}:[s_i,s_f]\rightarrow\mathbb{C}$, a path integral propagator that encodes `time-evolution' along $\mathcal{T}$ can be defined as follows:
	\begin{align}\label{kernelqmT2}
	\mathcal{G}^{\mathcal{T}}_{q}(q_f,z_f;q_i,z_i)= e^{i S^{\mathcal{T}}_{cl}(q_f,z_f;q_i,z_i)}\underbrace{\int \mathcal{D}y \exp\left(-\frac{i}{2}\int_{\mathcal{T}}dz \,y(z)\hat{O}y(z)\right)}_{\equiv \mathcal{F}^{\mathcal{T}}(z_f,z_i)}
	\end{align} 
	where, $z_i=\mathcal{T}(s_i)$ and $z_f=\mathcal{T}(s_f)$, $S^{\mathcal{T}}_{cl}$ is the classical action with time along $\mathcal{T}$, and the functions $y:\mathcal{T}\rightarrow\mathbb{C}$ vanish at $z_i$ and $z_f$. A straight forward extension of the arguments in \ref{setup} can be used to show that \ref{detandggen} generalizes to
	\begin{align}\label{detandggent2}
	\frac{\det_{\mathrm{\mathcal{T}}}(\hat{O}_2)}{\det_{\mathrm{\mathcal{T}}}(\hat{O}_1)}=\frac{g_2(z_f;z_i)\dot{g}_1(z_i;z_i)m_1(z_i)}{g_1(z_f;z_i)\dot{g}_2(z_i;z_i)m_2(z_i)}
	\end{align} 
	where, we have to use the prescription in \ref{prescription} to define the operators $\hat{O}_2$ and $\hat{O}_1$ in $\mathcal{T}$.

	\section{Reduction of order}\label{redorder}
	Consider the following differential equation.
	\begin{align}\label{difeq}
	\frac{d}{dt}\left(m\dot{q}\right)+\omega^2(t)q=0
	\end{align}
	Let $\xi$ be a solution of this equation. We seek for an independent solution of \ref{difeq} of the form $\tilde{\xi}(t)=\xi(t)g(t)$. In order to find $g(t)$, we substitute our ansatz into \ref{difeq} to get
	\begin{align}
	\frac{d}{dt}\left(\dot{g}m\xi^2\right)=0
	\end{align}
	We can easily integrate this equation to find $g$ to be
	\begin{align}
	g(t)=\textrm{constant}\times\int_{t_0}^{t}\frac{dt'}{m(t')\xi^2(t')}
	\end{align} 
	Therefore, the most general solution of \ref{difeq} can be written as
	\begin{align}
	q(t)=A\xi(t)\left(1+B\int_{t_0}^{t}\frac{dt'}{m(t')\xi^2(t')}\right)
	\end{align}
	where, $A$ and $B$ are constants, to be determined by the boundary/initial conditions.
	
	\subsection{Out mode in terms of in mode}\label{app1}
	Consider the following integral
	\begin{align}
	\int_{t}^{\infty}\frac{f_{in}^{*}(t;C)dt''}{m(C(t''))f_{in}^{*2}(t'';C)}
	\end{align}
	From our discussion, we see that this is solutions of \ref{difeq}. The boundary condition satisfied by this function can be understood by looking at the $t\rightarrow\infty$ limit.
	\begin{align}
	\int_{t}^{\infty}&\frac{f_{in}^{*}(t;C)dt''}{m(C(t''))f_{in}^{*2}(t'';C)}\\\nonumber
	&\approx \left(\alpha^*(t_f)e^{i\rho_f}\frac{e^{i\omega_f(t-t_f)}}{\sqrt{2m_f\omega_f}}\right)\left(\frac{2m_f\omega_f}{(\alpha^*(t_f))^2e^{2i\rho_f}}\right)\int_{t}^{\infty}\frac{e^{-2i\omega_f(t'-t_f)}}{m_f}dt'\\
	&\approx \frac{1}{i\alpha^*(t_f)}\left(e^{-i\rho_f}\frac{e^{-i\omega_f(t-t_f)}}{\sqrt{2\omega_f m_f}}\right)
	\end{align}
	Hence, we see that
	\begin{align}\label{result1}
	\int_{t}^{\infty}\frac{f_{in}^{*}(t;C)dt''}{m(C(t''))f_{in}^{*2}(t'';C)}    &=\frac{-i}{\alpha^*(t_f)}f_{out}(t;C)
	\end{align}
	Differentiating with respect to $t$ on both sides we also get
	\begin{align}\label{result2}
	\frac{-i}{\alpha^*(t_f)}\dot{f}_{out}(t;C)=\dot{f_{in}^{*}}(t;C)\int_{t}^{\infty}\frac{dt''}{m(C(t''))f_{in}^{*2}(t'';C)}-\frac{1}{m(C(t))f_{in}^{*}(t;C)}
	\end{align}
	
	A similar result can be obtained by replacing the time integrals with that along the contour $\mathcal{T}_2$.
	\begin{align}\label{res1a}
	\int_{z}^{(-\infty e^{-i\epsilon})}\frac{f_{in}^{*}(z;C)dz''}{m(C(z''))f_{in}^{*2}(z'';C)}=-if_{in}(z),
	\end{align}
	as well as,
	\begin{align}\label{res2a}
	D_z{f}_{in}^{*}(z;C)\int_{z}^{(-\infty e^{-i\epsilon})}\frac{dz''}{m(C(z''))f_{in}^{*2}(z'';C)}-\frac{1}{m(C(z))f_{in}^{*}(z;C)}&=-iD_z{f}_{in}(z;C)
	\end{align}
	
	\section{Derivation of functional derivative}\label{funcderiv}
	We shall first derive the the functional derivative of $\log[\textrm{det}_{\mathcal{T}_1}(\hat{O}[C])]$. The techniques can be easily generalized for a general contour $\mathcal{T}$.
	
	It is convenient to define
	\begin{align}\label{defines}
	\frac{\delta f_{in}^{*}(t;C)}{f_{in}^{*}(t;C)}=s(t;C)
	\end{align} 
	The differential equation satisfied by $s$, to first order, is given by \cite{Singh:2013pxf}
	\begin{align}\label{diffeqfors}
	\ddot{s}+A(t;C)\dot{s}+B(t;C)=0
	\end{align}
	where,
	\begin{align}
	A(t;C)&=\frac{d}{dt}\log\left[ m(C) f_{in}^{*2}(t;C)\right]\\
	B(t;C)&=\delta\omega^2(C)+\frac{\partial_t{f_{in}^{*}}(t;C)}{f_{in}^{*}(t;C)}\delta\left(\frac{\dot{m}}{m}\right)\\
	&=\left[\frac{\delta\omega^2(C)}{\delta C}+\frac{\partial_t{f_{in}^{*}}(t;C)}{f_{in}^{*}(t;C)}\frac{\delta\mu(C)}{\delta C}\right]\delta C+\frac{\partial_t{f_{in}^{*}}(t;C)}{f_{in}^{*}(t;C)}\frac{\delta\mu(C)}{\delta \dot{C}}\delta \dot{C}
	\end{align}
	where, $\mu(C)\equiv\dot{m}/m=\dot{C}(\partial_{C}m)/m$. The initial conditions are
	\begin{align}
	\lim_{t\rightarrow -\infty}s(e^{i\epsilon}t)=\lim_{t\rightarrow -\infty}\dot{s}(e^{i\epsilon}t)=0
	\end{align}
	Such a solution to \ref{diffeqfors} can be explicitly found (see \ref{funcderiv}) and is given by
	\begin{align}
	s(t)&=\int_{-\infty e^{i\epsilon}}^{t} dt''e^{-\gamma(t'';C)}\int_{-\infty e^{i\epsilon}}^{t''}dt'e^{\gamma(t';C)}B(t';C)\\
	&=\int_{-\infty e^{i\epsilon}}^{t}dt''e^{-\gamma(t'';C)}\int_{-\infty e^{i\epsilon}}^{t''}dt'e^{\gamma(t';C)}\left\{\left[\frac{\delta\omega^2(C)}{\delta C}\bigg |_{t'}+\frac{\partial_t{f_{in}^{*}}(t';C)}{f_{in}^{*}(t';C)}\frac{\delta\mu(C)}{\delta C}\bigg |_{t'}\right]\delta C(t')\right.\\\nonumber
	&+\left.\left[\frac{\partial_t{f_{in}^{*}}(t';C)}{f_{in}^{*}(t';C)}\frac{\delta\mu(C)}{\delta \dot{C}}\bigg |_{t'}\right]\delta \dot{C}(t')\right\}
	\end{align}   
	
	Let us define $\dot{s}=u$, so that \ref{diffeqfors} becomes,
	\begin{align}
	\dot{u}+Au+B=0
	\end{align}
	With $\gamma(t;C)=\log(m f_{in}^{*2})$, this equation can be rewritten as
	\begin{align}
	\frac{d}{dt}\left(e^{\gamma}u\right)+e^{\gamma}B=0
	\end{align}
	which can be easily integrated to get
	\begin{align}\label{solutionu}
	u(t)=e^{-\gamma(t;C)}\int_{-\infty e^{i\epsilon}}^{t}dt'e^{\gamma(t';C)}B(t';C)
	\end{align}
	where, we have used the condition $\dot{s}(-\infty e^{i\epsilon})=0$. One more of integration of \ref{solutionu} gives
	\begin{align}
	s(t)&=\int_{-\infty e^{i\epsilon}}^{t} dt''e^{-\gamma(t'';C)}\int_{-\infty e^{i\epsilon}}^{t''}dt'e^{\gamma(t';C)}B(t';C)\\
	&=\int_{-\infty e^{i\epsilon}}^{t}dt''e^{-\gamma(t'';C)}\int_{-\infty e^{i\epsilon}}^{t''}dt'e^{\gamma(t';C)}\left\{\left[\frac{\delta\omega^2(C)}{\delta C}\bigg |_{t'}+\frac{\partial_t{f_{in}^{*}}(t';C)}{f_{in}^{*}(t';C)}\frac{\delta\mu(C)}{\delta C}\bigg |_{t'}\right]\delta C(t')\right.\\\nonumber
	&+\left.\left[\frac{\partial_t{f_{in}^{*}}(t';C)}{f_{in}^{*}(t';C)}\frac{\delta\mu(C)}{\delta \dot{C}}\bigg |_{t'}\right]\delta \dot{C}(t')\right\}
	\end{align}
	where, we have used $s(-\infty e^{i\epsilon})=0$. 
	
	From \ref{detintermsoffin}, \ref{claim2} and \ref{defines}, we get
	\begin{align}
	&\delta\log[\textrm{det}_{\mathcal{T}_1}(\hat{O}[C])]=\int_{-\infty e^{i\epsilon}}^{\infty e^{i\epsilon}}dt''\int_{-\infty e^{i\epsilon}}^{t''}dt'\left[\frac{m(C(t'))f_{in}^{*2}(t';C)}{m(C(t''))f_{in}^{*2}(t'';C)}\right]\\ \nonumber
	&\times\bigg\{\left[\frac{\delta\omega^2(C)}{\delta C}\bigg |_{t'}+\frac{\partial_t{f_{in}^{*}}(t';C)}{f_{in}^{*}(t';C)}\frac{\delta\mu(C)}{\delta C}\bigg |_{t'}\right]\delta C(t')
	+\left[\frac{\partial_t{f_{in}^{*}}(t';C)}{f_{in}^{*}(t';C)}\frac{\delta\mu(C)}{\delta \dot{C}}\bigg |_{t'}\right]\delta \dot{C}(t')\bigg\}
	\end{align}
	After doing an integration-by-parts on the term with $\delta\dot{C}$, the last equation simplifies to
	\begin{align}
	&\delta\log[\textrm{det}_{\mathcal{T}_1}(\hat{O}[C])]=\\\nonumber
	&\int_{-\infty e^{i\epsilon}}^{\infty e^{i\epsilon}}dt''\int_{-\infty e^{i\epsilon}}^{t''}dt'\left[\frac{m(C(t'))f_{in}^{*2}(t';C)}{m(C(t''))f_{in}^{*2}(t'';C)}\right]\\ \nonumber
	&\times\left\{\left[\frac{\delta\omega^2(C)}{\delta C}\bigg |_{t'}+\frac{\partial_t{f_{in}^{*}}(t';C)}{f_{in}^{*}(t';C)}\frac{\delta\mu(C)}{\delta C}\bigg |_{t'}\right]\delta C(t')-\frac{d}{dt'}\left[\frac{\partial_t{f_{in}^{*}}(t';C)}{f_{in}^{*}(t';C)}\frac{\delta\mu(C)}{\delta \dot{C}}\bigg |_{t'}\right]\delta C(t')\right\}\\ \nonumber
	&+\int_{-\infty e^{i\epsilon}}^{\infty e^{i\epsilon}}\frac{dt''}{m(C(t''))f_{in}^{*2}(t'';C)}\left[\frac{\partial_t{f_{in}^{*}}(t'';C)}{f_{in}^{*}(t'';C)}\frac{\delta\mu(C)}{\delta \dot{C}}\bigg |_{t''}\right]\delta C(t'')
	\end{align}
	Therefore, the functional derivative of $\delta\log[\textrm{det}_{\mathcal{T}_1}(\hat{O}[C])]$ is given by
	\begin{align}\label{deldetdelc1app}
	i\frac{\delta\log[\textrm{det}_{\mathcal{T}_1}(\hat{O}[C])]}{\delta C(t)}&=\int_{-\infty e^{i\epsilon}}^{\infty}dt''\int_{-\infty e^{i\epsilon}}^{t''}dt'\left[\frac{m(C(t'))f_{in}^{*2}(t';C)}{m(C(t''))f_{in}^{*2}(t'';C)}\right]\\\nonumber
	&\times\left\{\left[\frac{\delta\omega^2(C)}{\delta C}\bigg |_{t'}+\frac{\partial_t{f_{in}^{*}}(t';C)}{f_{in}^{*}(t';C)}\frac{\delta\mu(C)}{\delta C}\bigg |_{t'}\right]\delta (t-t')\right.\\\nonumber
	&\left.-\frac{d}{dt'}\left[\frac{\partial_t{f_{in}^{*}}(t';C)}{f_{in}^{*}(t';C)}\frac{\delta\mu(C)}{\delta \dot{C}}\bigg |_{t'}\right]\delta(t-t')\right\}\\\nonumber
	&+\int_{-\infty e^{i\epsilon}}^{\infty e^{i\epsilon}}\frac{dt''}{m(C(t''))f_{in}^{*2}(t'';C)}\left[\frac{\partial_t{f_{in}^{*}}(t'';C)}{f_{in}^{*}(t'';C)}\frac{\delta\mu(C)}{\delta \dot{C}}\bigg |_{t''}\right]\delta(t-t'')
	\end{align}
	Using the following results (see \ref{result1} and \ref{result2} of \ref{app1})
	\begin{align}\label{intonebyfapp}
	\int_{t}^{\infty e^{i\epsilon}}\frac{f_{in}^{*}(t;C)dt''}{m(C(t''))f_{in}^{*2}(t'';C)}    &=\frac{-i}{\alpha^*(t_f)}f_{out}(t;C)\\
	\dot{f}_{in}^{*}(t;C)\int_{t}^{\infty e^{i\epsilon}}\frac{dt''}{m(C(t''))f_{in}^{*2}(t'';C)}-\frac{1}{m(C(t))f_{in}^{*}(t;C)}&=\frac{-i}{\alpha^*(t_f)}\dot{f}_{out}(t;C)
	\end{align}
	the right hand side of \ref{deldetdelc1app} can be simplified to get
	\begin{align}\label{deldetdelcapp}
	i\frac{\delta\log[\textrm{det}_{\mathcal{T}_1}(\hat{O}[C])]}{\delta C(t)}&=\frac{-1}{\alpha^*(t_f)}\left[\left(\partial_{C}m\right)\dot{f}_{out}(t;C)\dot{f}^*_{in}(t;C)-\partial_C(m\omega^2)f_{out}(t;C)f^*_{in}(t;C)\right]
	\end{align}
	After a bit of algebra (see \ref{funcderiv} for details), this expression can be further simplified to give:
	\begin{align}\label{deldetdelc}
	i\frac{\delta\log[\textrm{det}_{\mathcal{T}_1}(\hat{O}[C])]}{\delta C(t)}&=\frac{-1}{\alpha^*(t_f)}\left[\left(\partial_{C}m\right)\dot{f}_{out}(t;C)\dot{f}^*_{in}(t;C)-\partial_C(m\omega^2)f_{out}(t;C)f^*_{in}(t;C)\right]
	\end{align}
	The following results (see \ref{inoutcorrelator} for details)
	that relates the terms in \ref{deldetdelc} with `in-out' matrix elements can be used to further simplify the expression for the functional derivative:
	\begin{align}
	\frac{\braket{\textrm{out}|q^2(t)|\textrm{in}}}{\braket{\textrm{out}|\textrm{in}}}&=\frac{f_{out}(t;C)f_{in}^{*}(t;C)}{\alpha^*(t_f)};&&\frac{\braket{\textrm{out}|p^2(t)|\textrm{in}}}{\braket{\textrm{out}|\textrm{in}}}=\frac{m^2\dot{f}_{out}(t;C)\dot{f}_{in}^*(t;C)}{\alpha^*(t_f)}
	\end{align}
	A direct substitution of these results in \ref{deldetdelc} yields:
	\begin{align}
	\frac{\delta\log[\textrm{det}_{\mathcal{T}_1}(\hat{O}[C])]}{\delta C(t)}&=\partial_{C}(m^{-1})\frac{\braket{\textrm{out}|p^2(t)|\textrm{in}}}{\braket{\textrm{out}|\textrm{in}}}+\partial_C(m\omega^2)\frac{\braket{\textrm{out}|q^2(t)|\textrm{in}}}{\braket{\textrm{out}|\textrm{in}}}
	\end{align}
	
	A similar analysis can be done for the case of contour $\mathcal{T}$. We obtain
	\begin{align}\label{deldetdelczapp}
	i\frac{\delta\log[\textrm{det}_{\mathcal{T}}(\hat{O}[C])]}{\delta C(z)}&=\int_{\mathcal{T}_2\vert_{z}}dz''\int_{\mathcal{T}\vert_{z''}}dz'\left[\frac{m(C(z'))f_{\sigma}^{*2}(z';C)}{m(C(z''))f_{\sigma}^{*2}(z'';C)}\right]\\\nonumber
	&\times\left\{\left[\frac{\delta\omega^2(C)}{\delta C}\bigg |_{z'}+\frac{D_z{f_{\sigma}^{*}}(z';C)}{f_{\sigma}^{*}(z';C)}\frac{\delta\mu(C)}{\delta C}\bigg |_{z'}\right]\delta(z-z')\right.\\\nonumber
	&\left.-D_{z'}\left[\frac{D_z{f_{\sigma}^{*}}(z';C)}{f_{\sigma}^{*}(z';C)}\frac{\delta\mu(C)}{\delta \dot{C}}\bigg |_{z'}\right]\delta(z-z')\right\}\\\nonumber
	&+\int_{\mathcal{T}}\frac{dz''}{m(C(z''))f_{\sigma}^{*2}(z'';C)}\left[\frac{D_z{f_{\sigma}^{*}}(z'';C)}{f_{\sigma}^{*}(z'';C)}\frac{\delta\mu(C)}{\delta \dot{C}}\bigg |_{z''}\right]\delta(z-z'')
	\end{align}
	The special of this equation for $\mathcal{T}=\mathcal{T}_2$ gives us \ref{deldetdelcz}.
	Using \ref{res1a} and \ref{res2a} in \ref{deldetdelcz}, this special case of the functional derivative can be simplified to
	\begin{align}\label{deldetdel4app}
	i\frac{\delta\log[\det(\hat{O}[C])]}{\delta C(z)}&=-\left[\left(\partial_{C}m\right)\dot{f}_{in}(z;C)D_z{f}^*_{in}(z;C)-\partial_C(m\omega^2)f_{in}(z;C)f^*_{in}(z;C)\right]
	\end{align}

	\section{The `in-out' correlator}\label{inoutcorrelator}
	In this section we will derive the `in-out' correlator for the $q$-system using the Heisenberg picture. The time evolution of any observable of the $q$-system can be constructed out of the time dependent creation and annihilation operators $a$ and $a^{\dagger}$. The quantum dynamics of the system is then described by the following two equations.
	\begin{align}\label{evolvea}
	a(t)&=\alpha(t)a_{i}+\beta^*(t)a_{i}^{\dagger}\\ \label{evolveadag}
	a^{\dagger}(t)&=\beta(t)a_{i}+\alpha^*(t)a_{i}^{\dagger}
	\end{align}
	where, $a_{i}\equiv a(t_i)$ and $a_{i}^{\dagger}\equiv a^{\dagger}(t_i)$. The evolution of position operator can be written as
	\begin{align}\label{qin}
	q(t)=f_{in}(t)a_{i}+f^*_{in}(t)a_{i}^{\dagger}
	\end{align}
	where, $f_{in}^{*}$ are the `in-modes'. Alternatively,
	\begin{align}\label{qout}
	q(t)=f_{out}(t)a_{f}+f^*_{out}(t)a_{f}^{\dagger}
	\end{align}
	where, $a_{f}\equiv a(t_f)$ and $a_{f}^{\dagger}\equiv a^{\dagger}(t_f)$. Consider the following correlator.
	\begin{align}
	\frac{\braket{\textrm{out}|q(t_2)q(t_1)|\textrm{in}}}{\braket{\textrm{out}|\textrm{in}}}
	\end{align}
	It is convenient to expand $q(t_2)$ using \ref{qout} and $q(t_1)$ using \ref{qin} so that,
	\begin{align}
	\frac{\braket{\textrm{out}|q(t_2)q(t_1)|\textrm{in}}}{\braket{\textrm{out}|\textrm{in}}}&=\frac{\braket{\textrm{out}|\left\{f_{out}(t_2)a_{f}+f^*_{out}(t_2)a_{f}^{\dagger}\right\}\left\{f_{in}(t_1)a_{i}+f^*_{in}(t_1)a_{i}^{\dagger}\right\}|\textrm{in}}}{\braket{\textrm{out}|\textrm{in}}}\\
	&=f_{out}(t_2)f_{in}^{*}(t_1)\frac{\braket{\textrm{out}|a_fa_i^{\dagger}|\textrm{in}}}{\braket{\textrm{out}|\textrm{in}}}
	\end{align}
	Using \ref{evolveadag} we can rewrite the matrix element in the last equation as
	\begin{align}
	\frac{\braket{\textrm{out}|a_fa_i^{\dagger}|\textrm{in}}}{\braket{\textrm{out}|\textrm{in}}}&=\frac{\braket{\textrm{out}|a_f\left\{\alpha(t_f)a_{f}^{\dagger}-\beta(t_f)a_{f}\right\}|\textrm{in}}}{\braket{\textrm{out}|\textrm{in}}}\\
	&=\alpha(t_f)\frac{\braket{\textrm{out}|a_fa^{\dagger}_f|\textrm{in}}}{\braket{\textrm{out}|\textrm{in}}}-\beta(t_f)\frac{\braket{\textrm{out}|a_f^2|\textrm{in}}}{\braket{\textrm{out}|\textrm{in}}}\\
	&=\alpha(t_f)-\beta(t_f)\frac{\beta^*(t_f)}{\alpha^*(t_f)}\\
	&=\frac{1}{\alpha^*(t_f)}
	\end{align}
	where we have used the following result
	\begin{align}
	\frac{\braket{\textrm{out}|a_f^2|\textrm{in}}}{\braket{\textrm{out}|\textrm{in}}}=\frac{\beta^*(t_f)}{\alpha^*(t_f)}
	\end{align}
	which follows easily from the expansion of $\ket{\textrm{m}}$ in terms of the complete set of `in-states' $\{\ket{n;\textrm{in}}\}$. Therefore, the `in-out' correlator becomes
	\begin{align}
	\frac{\braket{\textrm{out}|q(t_2)q(t_1)|\textrm{in}}}{\braket{\textrm{out}|\textrm{in}}}&=\frac{f_{out}(t_2)f_{in}^{*}(t_1)}{\alpha^*(t_f)}
	\end{align}
	The normalized `in-out' matrix element of $q^2$ is obtained by putting $t_2=t_1=t$ in the above equation.
	\begin{align}\label{result3}
	\frac{\braket{\textrm{out}|q^2(t)|\textrm{in}}}{\braket{\textrm{out}|\textrm{in}}}&=\frac{f_{out}(t)f_{in}^{*}(t)}{\alpha^*(t_f)}
	\end{align}
	Similarly, one can show that
	\begin{align}\label{result4}
	\frac{\braket{\textrm{out}|p^2(t)|\textrm{in}}}{\braket{\textrm{out}|\textrm{in}}}=\frac{m^2\dot{f}_{out}(t)\dot{f}_{in}^*(t)}{\alpha^*(t_f)}
	\end{align}

\end{document}